\documentclass[aps,pra,epsfigure,twocolumn,showpacs]{revtex4}
\usepackage{dcolumn}    
\usepackage{bm} 
\usepackage{graphicx}
\usepackage{amsmath}    
\usepackage{latexsym}
\usepackage{amsfonts}   
\usepackage{amssymb}
\usepackage{array}      
\usepackage{epsfig}
\usepackage{epstopdf}

\newcommand{\ket}[1]{\left\vert#1\right\rangle}

\newcommand{\bra}[1]{\left\langle#1\right\vert}

\begin{document}
\title{Structural change in multipartite entanglement sharing: a random matrix approach}
\author{G. Gennaro,$^{1*}$ S. Campbell,$^{2*}$ M. Paternostro,$^2$ and G. M. Palma,$^3$}
\affiliation{$^1$Dipartimento di Scienze Fisiche ed Astronomiche, 
Universita' degli Studi di Palermo, via Archirafi 36, I-90123 Palermo, Italy\\
$^2$ School of Mathematics and Physics. Queen's University Belfast, BT7 1NN, United Kingdom\\
$^3$ NEST-CNR-INFM \& Dipartimento di Scienze Fisiche ed Astronomiche, 
Universita' degli Studi di Palermo, via Archirafi 36, I-90123 Palermo, Italy}

\begin{abstract}
We study the typical entanglement properties of a system comprising two independent qubit environments interacting via a shuttling ancilla. The initial preparation of the environments is modelled using random-matrix techniques. The entanglement measure used in our study is then averaged over many histories of randomly prepared environmental states. Under a Heisenberg interaction model, the average entanglement between the ancilla and one of the environments remains constant, regardless of the preparation of the latter and the details of the interaction. We also show that, upon suitable kinematic and dynamical changes in the ancilla-environment subsystems, the entanglement-sharing structure undergoes abrupt modifications associated with a change in the multipartite entanglement class of the overall system's state. These results are invariant with respect to the randomized initial state of the environments.
\end{abstract}
\date{\today}
\maketitle


Open-system dynamics involving environments comprising only a finite number of elements have been proven a valuable arena for the study of interesting physical phenomena such as quantum chaos~\cite{mahler}, quantum thermodynamics~\cite{thermod}, entanglement and relaxation~\cite{giulianomassimo,giuseppegiulianomassimo}. In most of these studies, the use of random-matrix theory has proven technically very advantageous in modelling random {\it collisions} between parts of the overall system at hand. Random matrices have been helpful in dealing with many tasks of quantum information processing, including quantum data hiding, quantum state distinction, superdense coding and noise estimation~\cite{applicazionirandom}. Very recently, random matrices have found extensive application in the characterization of Markovian decoherence~\cite{pineda}. 

In this paper we unveil a further interesting situation where the theory of random matrices finds fertile applications: We study the typical amount of entanglement that can be set in a multipartite system comprising two environments of arbitrary (finite) size and a shuttling two-level ancilla that bridges their cross-talking (Fig. 1). Differently from Refs.~\cite{giulianomassimo,giuseppegiulianomassimo,pineda,gennaro}, random matrices are used in order to model the initial preparation of the environments. These interact with the shuttling ancilla via a Hamiltonian model having pre-determined interaction strength. This allows us to investigate on the {\it typical} ancilla-environment as well as all-environment degree of entanglement simply by averaging the values corresponding to many hystories of random initial preparations. We point out the sensitivity of the entanglement-sharing structure on the kinematic and dynamical aspects of the interactions and its independence from the random preparation of the environments. In particular, we show that upon tuning of the coupling Hamiltonian regulating the ancilla-environment interactions, an abrupt transition between two inequivalent classes of multipartite entanglement is achieved. The dimensions of the environments and the number of interactions enter preponderantly into the determination of the entanglement-sharing structure, as we quantitatively reveal.

It is important to remark that our model does not allow for phenomena of thermalization or homogeneization typical of effectively Markovian {\it reservoirs} such as those considered in Refs.~\cite{omogeneizzazione}. In fact, the evolution here at hand is profoundly different from a forgetful open dynamics comprising environments which are weakly perturbed by the interactions with the ancilla. Memory effects are clearly seen in our analysis, which reveals how the state of each environment is deeply affected by its coupling to the shuttling ancilla. As we do not impose restrictions to the strength of each interaction, also the Born approximation does not hold in our analysis. 

The remainder of this paper is organized as follows. In Sec.~\ref{onereg} we address the case of a single environment interacting with the ancilla $A$. Through numerical simulations supported by a clear analytical study we show that when the ancilla interacts repeatedly with the same environmental qubit, the degree of entanglement is invariant to both the random initial preparation and dimension of the environment. Sec.~\ref{tworeg} extends our study to the case of two multi-qubit environments mutually connected by the bouncing ancilla. $A$ repeatedly interacts ({\it collides}, as we often say in this paper) with two specific qubits, each beloning to the respective environment. In this case, the amount of entanglement that can be set is a delicate trade-off between both the dimensions of the two environments and the order of the ancilla-environment interaction. This scenario allows us to address the main points of our study. First we show that, by tuning the form of the coupling Hamiltonian, the system undergoes an entanglement-sharing transition between two non-equivalent classes of multipartite entanglement. As no time is explicitly involved, we refer to this effect as a ``kinematic" transition. We then study how repeated sequential collisions are able to affect the bipartite and tripartite entanglement in a way so as to mark a second, more dynamical entanglement-sharing transition. Sec.~\ref{conclusions} summarizes our findings and opens up perspectives for future work. Finally, an appendix is devoted to the technical description of the way random unitaries are built in this work. 

\section{Single-environment case: Propaedeutic results}
\label{onereg}

In this Section we study the single environment-ancilla system, which serves a useful case for the discussion of a few preparatory results that will then be reprised when the two-enviroment situation is studied.
\begin{figure}[b]
\psfig{figure=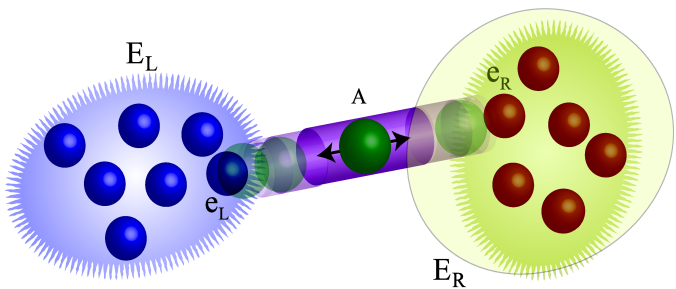,width=8.9cm,height=3.8cm}
\caption{(Color online).  Sketch of the situation considered. An ancillary shuttle A, modeled as a two-level system, interacts with one of the elements of two spin environments, $E_{L,R}$, which are remotely located. The initial preparation of each environment state is random. Upon control over the details of the interactions and the order of the collisions with A, effective manipulation of the entanglement-sharing structure of the multipartite system comprising environments and shuttle is achieved. Environment $E_R$ is shadowed in order to indicate that we consider both the single-environment and two-environment cases. }
\label{model}
\end{figure}
Let us consider a pure, initially separable state of an $n$-qubit environment $E=\{e_1,e_2,..,e_n\}$. As a reference-state for our analysis, we assume that all the qubits are initially in their respective ground state so that we have
\begin{equation}
\ket{\psi}_E=\bigotimes^n_{k=1}\ket{0}_{k}.
\end{equation}
We then introduce an ancillary two-level system, $A$, with which $E$ interacts. The ancilla is assumed to be prepared in its ground state $\ket{0}_A$ and it is meant to {\it collide} with a specific element of $E$, which we label $e$. The assumption on the preparation of A is simply matter of convenience as far as pure states are taken. Moreover, we will show that for $\dim{(E)}=1$ any state can be considered for the ancilla, including mixed states. The Hamiltonian modeling such an interaction is taken to be of the anisotropic Heisenberg form 
\begin{equation}
\label{modelH}
\hat{H}_{Ae}=\sum^3_{i=1} J_{i}\hat{\sigma}_{i,A} \otimes\hat{\sigma}_{i,e}
\end{equation}
where $J_i$ is an interaction strength and $\hat{\sigma}_{i,A(e)}$'s are the Pauli spin operators of the $A$ ($e$) qubit. We have used the labelling $\hat{\sigma}_1=\hat{\sigma}_x$, $\hat{\sigma}_2=\hat{\sigma}_y$ and $\hat{\sigma}_3=\hat{\sigma}_z$. Our task is to investigate typical entanglement properties of the $e-A$ system and we thus aim at removing any dependence of our analysis from the state of the environment $E$. In order to achieve this we proceed as follows: we prepare $E$ in the state resulting from the application of a random $n$-partite quantum gate~\cite{NC} constructed using the random unitary matrix $\hat{U}_r$ that we uniformly draw from the normalized Haar measure of the unitary group~\cite{hurwitz,kus1994,kus1991}. In the Appendix we provide the technical details for the parameterization of such random unitaries, which are the building block of our numerical calulations. We then let the $e-A$ system evolve as $\varrho=\hat{U}_{h}\hat{U}_r \rho\hat{U}^{\dagger}_r\hat{U}_{h}^{\dagger}$, where $\hat{U}_h= e^{-i\hat{H}_{Ae} t}$ and $\rho= \ket{\psi}_E\bra{\psi}\otimes \ket{0}_A\bra{0}$ and determine the associated amount of entanglement. This is then statistically averaged over a large ensemble of random initial preparations so as to remove the dependence on the initial environmental state. 

 We now anticipate a quantitative result achieved via our numerical calculations, which will then be justified in a fully analytical way: for an isotropic Heisenberg interaction defined by taking $J_{x,y,z} = J$ in Eq.~(\ref{modelH}), the entanglement set between $A$ and $e$ only depends on the dimensionless interaction time $Jt$ and [as far as the ancilla is prepared in a pure state] is insensitive to the dimension of $E$, $\dim(E)$, and the preparation of $A$. One may think that this result is only due to the fact that $A$ collides with a single environmental qubit, thus reducing our problem to a two-qubit interaction. However, this is definitely not the case. In fact, in general, the random evolution within the environment, encompassed by the random unitary matrices $U_r$'s, creates a multipartite entangled state. Although the ancilla $A$ physically collides with $e$ only, the state of the latter is crucially affected by the entanglement-sharing structure within $E$, which is highly non trivial. In Ref.~\cite{kendon}, for instance, it was shown that the behavior of bipartite entanglement shared by any two elements of a multipartite register $E$ is a decreasing function of $\dim(E)$. Intuition would then lead one to believe that the entanglement set between the interacting qubits would also depend on $\dim(E)$~\cite{referee2}. Here we show that such a dependence is not in order. In order to understand this counter-intutive behavior, we take a two-step approach. First, by studying the case of $\dim(E)=1$, we provide an intuition of the reasons behind the dependence of the typical amount of $A-e$ entanglement on the sole interaction strength $J$. In this simple and yet useful case, a random unitary operation built according to the parameterization given in the Appendix is represented by the $2\times{2}$ matrix 
 \begin{equation}
\label{euler}
U_r=e^{i \alpha}
\left(\begin{array}{cc} 
\cos \phi e^{i\psi} & \sin \phi e^{i \chi}\\
-\sin \phi e^{-i \chi} & \cos \phi e^{-i\psi}
\end{array}\right),
\end{equation}
where the angles $\alpha,\chi,\psi$ ($\phi$) are picked up from the range $[0,2\pi]$ ($[0,\pi/2]$) uniformly with respect to the normalized Haar measure $dU_r=(8\pi^3)^{-1}\sin(2\phi)d\phi{d}\psi{d}\chi{d}\alpha$~\cite{kus1994,buda,nota}. In order to evaluate the entanglement set within the system, we use Wootters' concurrence~\cite{CONC}. For a general bipartite pure or mixed state described by the density matrix $\varrho$, concurrence is given by
\begin{equation}
{\cal C}= \max [0,\sqrt{\lambda_1}-\sum^4_{k=2}\sqrt{\lambda_k}]
\end{equation}
where $\lambda_1\ge\lambda_j~(j=2,3,4)$ are the eigenvalues of $\rho\,(\sigma_2\otimes\sigma_2)\, \rho^*\,(\sigma_2\otimes\sigma_2)$. When $A$ is prepared in its ground state and upon evolution of the $e-A$ system as $\varrho=\hat{U}_{h}\hat{U}_r \rho\hat{U}^{\dagger}_r\hat{U}_{h}^{\dagger}$, we get the density matrix
\begin{widetext}
\begin{equation}
\label{matricedensita}
\varrho\!=\!
\begin{pmatrix}
C^{2}(\phi) & -\frac{i}{2}e^{-2iJt+i(\psi+\chi)}S(2Jt) S(2\phi) & -\frac{1}{2}e^{-2iJt+i(\psi+\chi)} C(2Jt) S(2\phi) & 0\\
\frac{i}{2}e^{2iJt-i(\psi+\chi)}S(2Jt) S(2\phi) & S^{2}(\phi) S^{2}(2Jt) & -\frac{i}{2}S(4Jt)S^{2}(\phi) & 0\\
-\frac{1}{2}e^{2iJt-i(\psi+\chi)}C(2Jt)C(2\phi) & \frac{i}{2}S(4Jt)S^{2}(\phi) & C^{2}(2Jt)S^{2}(\phi) & 0\\
0 & 0 & 0 & 0
\end{pmatrix},
\end{equation}
\end{widetext}
with $C(x)=\cos(x)$ and $S(x)=\sin(x)$. It is then straightforward to see that the concurrence shared by $e$ and $A$ after the application of $\hat{U}_r$ and $\hat{U}_h$ results in the elegant expression
\begin{equation}
\label{first}
{\cal C}_{eA}=\sin^2(\phi)|\sin({4Jt})|.
\end{equation}
This shows that, given a specific preparation of the environment, the only parameters governing the entanglement are $\phi$ and $Jt$. The typical value of ${\cal C}_{eA}$ is obtained by averaging the above expression over any possible unitary matrix $U_r$ uniformly drawn according to the proper Haar measure. Explicitly, we have to calculate
\begin{equation}
\label{singleconc}
\begin{aligned}
\overline{\cal C}_{eA}&=\frac{1}{8\pi^{3}} \int^{\pi/2}_{0}\sin(2\phi)\,d{\phi}\int^{2\pi}_{0}d{\psi}\int^{2\pi}_{0}d\chi\int^{2\pi}_{0}d\alpha\, {\cal C}_{eA}\\
&= \frac{1}{2}\left|\sin(4J\tau)\right|,
\end{aligned}
\end{equation} 
which may seem a special result arising from the pure-state preparation of the ancilla. However, this is definitely not the case, as we now demonstrate. By starting with the mixed ancilla state 
\begin{equation}
\label{mixed}
\rho_A = 
\begin{pmatrix}
\rho_0&0\\ 
0&1-\rho_{0}  
\end{pmatrix},
~~\rho_0\in[0,1]
\end{equation}
and calculating the evolution of the $e-A$ system, one gets a density matrix that is an easy generalization of Eq.~(\ref{matricedensita}). By inspecting the eigenvalues of $\rho\,(\sigma_2\otimes\sigma_2)\, \rho^*\,(\sigma_2\otimes\sigma_2)$, whose explicit form is too lenghty to be reported here, one finds no dependence on $\chi,\,\psi$ or $\alpha$, in full analogy with the case of Eq.~(\ref{singleconc}).
The explicit calculation of concurrence leads us to the expression
\begin{equation}
{\cal C}_{eA} = 
\frac{1}{2}|[-1 + (2\rho_0-1)\cos(2\phi)]\sin(4Jt)|.
\end{equation} 
As $\int^{\pi/2}_0\cos(2\phi)\sin(2\phi)d\phi=0$, the average concurrence (calculated using the appropriate Haar measure, as done before) turns out to be identical to Eq.~(\ref{first}). The study can be straightforwardly generalised to the case of qubit $A$ being 
prepared in any coherent-superposition state, the only difficulty being a slightly more complicated expression for the concurrence corresponding to any set preparation of $E$. The message, however, is rather clear: regardless of 
the state into which the ancilla is prepared, the typical $e-A$ entanglement is simply set 
by the rescaled interaction time $Jt$. This is strictly valid only 
in the statistical sense: if specific instances of preparation of 
both $A$ and $e$ are taken, such an independence does not hold 
anymore. However, contrary to a naive expectation, the typical entanglement does not vanish. As we see later, this result is the key to understand what occurs in the two-environment case. 

We now approach the second step of our proof by studying the invariance of the $e-A$ entanglement with respect to $\dim(E)$ when $A$ is prepared in a pure state. For this task, we consider a simple extension of the previous case to a two-qubit environment $E=\{e_1,e_2\}$, initially prepared in a state described by
\begin{equation}
\begin{aligned}
\rho_{E}&=\frac{1}{4}(\openone_{e_1}\!\otimes\openone_{e_2}+\sum^3_{k=1}\beta_{k,e_2}\openone_{e_1}\!\otimes\hat{\sigma}_{k,e_2}
\\&+\sum^3_{k=1}\beta_{k,e_1}\hat{\sigma}_{k,e_1}\!\otimes\openone_{e_2}
+\sum^3_{k,l=1}\chi_{kl}\hat{\sigma}_{k,e_1}\!\otimes\hat{\sigma}_{l,e_2})
\end{aligned}
\end{equation}
with ${\bm \chi}$ the elements of the tensor accounting for the correlations between $e_1$ and $e_2$ and ${\bm \beta}_{e_j}~(j=1,2)$ the Bloch vector of qubit $e_j~(j=1,2)$~\cite{NC,massimo}. This form holds for both entangled and separable two-qubit states and is thus a formal description of an arbitrary preparation of $E$. 
By taking $e\equiv{e}_2$ (an arbitrary choice that does not affect the generality of our discussion) we follow the recipe for evolution described above. This time, before calculating the $e-A$ concurrence, we have to trace over $e_1$'s degrees of freedom. Through a tedious but otherwise straightforward calculation, we see that although the tripartite $E-A$ density matrix depends on ${\bm \chi}$ and ${\bm \beta}_{e_1}$, the reduced density matrix of the $e-A$ system only depends on ${\bm \beta}_{e_2}$. 
By properly averaging over any possible preparation of the environmental qubit $e_2$, we are led to a typical $e-A$ entanglement that is identical to Eq.~(\ref{first}). These considerations can be extended to $\dim(E)=n$, although the complexity of an analytical proof scales exponentially with $n$. However, our numerical calculations are in perfect agreement with the analytic conclusions, supporting the general validity of the arguments used here and thus demonstrating the claimed invariance. 

The relevance of this result and its significance become evident when one considers that random-matrix theory is widely believed to provide a good effective description of the state of a system in thermal equilibrium at a high temperature~\cite{facchi}. In such unfavorable conditions, the achievement of entanglement regardless of the initial state of the environment ({\it i.e.} its effective temperature) and  its robustness against the complexity of $E$'s structure and the initial (pure) state of $A$ is a remarkable feature. Our study reveals that through the use of a suitable interaction we can reliably create entanglement between a {\it clean} ancilla qubit and a mixed-state environment qubit. Furthermore, the invariance of the amount of generated entanglement with respect to the size of the environment strongly contrasts with the intuitive belief that ${\cal C}_{eA}$ would have followed a monogamy constraints similar to those responsible for the bipartite entanglement in an $n$-qubit system~\cite{monogamy,kendon}. Besides its intrinsic interest, such an invariance will be proven crucial for the understanding of the study conducted in Sec.~\ref{tworeg}

\section{Two-environment case: Entanglement-structure transition}
\label{tworeg}
We now turn our attention to the main scenario of our investigation, which consists of two independent and spatially separated environments, $E_R$ and $E_L$. This time, the ancilla $A$ shuttles between them, colliding always with the same qubit of each environment. Again, we are interested in typical values of entanglement and, in order to cancel any dependence on the initial preparation of the environments, we rely on a random-matrix approach. We first show that, in line with the results unveiled in Sec.~\ref{onereg}, one of the environments is entangled with $A$ with a constant typical degree, dependent solely on $Jt$. This is accompanied by a few other results related to the details of the environment-ancilla interactions. Most importantly, however, we demonstrate that the genuine tripartite entanglement established among $A$ and the environmental qubits involved in the collisions is subjected to a drastic change in its structure, depending on the shape of the interaction Hamiltonian and the number of collisions. The fact that these results refer to typical ({\it i.e.} statistically averaged) quantities, make them even more intriguing. Not only multipartite entanglement can be generated: through suitable interaction engineering we can actually efficiently dictate its form and nature.

\subsection{Quantitative analysis of entanglement}
\label{quantan}

\begin{figure}[t]
{\bf (a)}\hskip3cm{\bf (b)}
\center{\psfig{figure=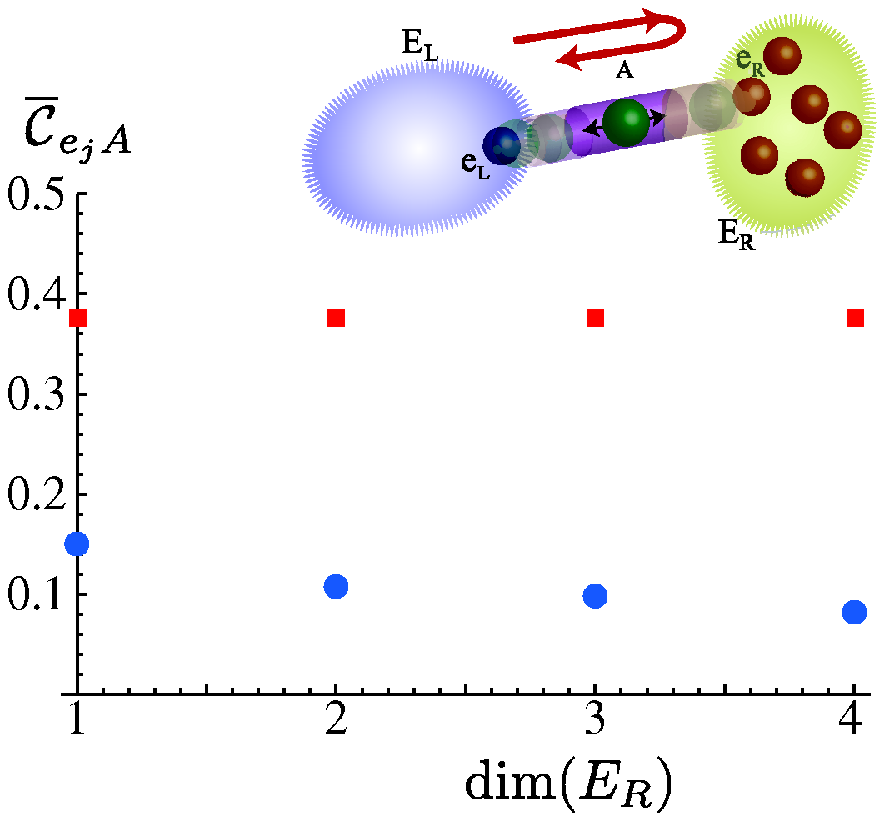,width=4.35cm,height=4.0cm}~~\psfig{figure=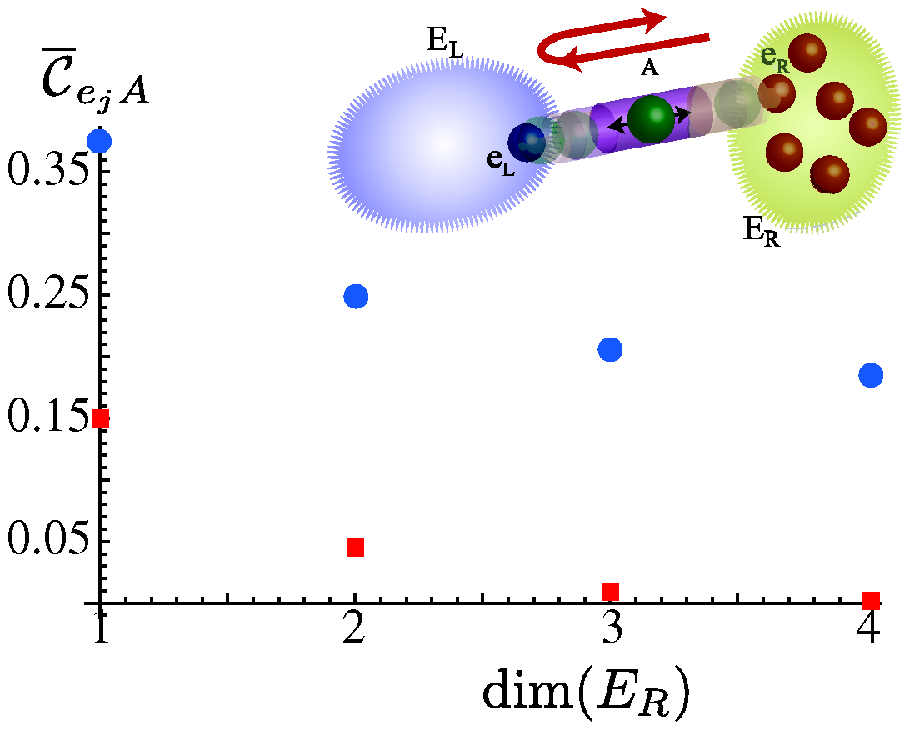,width=4.35cm,height=4.0cm}}
\caption{(Color online). Typical concurrence between the ancilla $A$ and the environment qubits $e_{L,R}$ for $\dim(E_L)=1$ and $\dim(E_R)=1,..,4$. In panel {\bf (a)}, $A$ interacts first with $e_R$ and then with $e_{L}$, for a rescaled interaction time $Jt = 1$ (see inset). The blue dots (red squares) show the average concurrence between $A$ and $e_R$ ($e_L$). In agreement with the analytic formula in Eq.~(\ref{first}), $\overline{\cal C}_{e_LA}=|\sin4|/2\simeq0.378$. {\bf (b)} Same as in panel {\bf (a)}, but with a reversed order of interactions. }
\label{newplot}
\end{figure}

Let us first consider the case where $\dim(E_L)=1$ (we label its only element as $\{e_L\}$), while $E_R$ consists of qubits $\{e_1,e_2,..,e_n\}$. If $A$ collides only once, first with $E_R$ and then $E_L$, we find that the degree of entanglement between the single-qubit environment $E_L$ and the ancilla remains constant against the number of qubits belonging to $E_R$, as show in Fig.~\ref{newplot} {\bf (a)}. We can easily understand this in light of the results discussed in the previous Section, where we have shown that the typical entanglement between $A$ and a single-qubit environment does not depend on the input state of the ancilla. As the collision with $E_R$, followed by the trace over its degrees of freedom, simply prepares $A$ in a mixed state, the invariance of $\overline{\cal C}_{e_LA}$ against $\dim(E_R)$ is clear: quantitatively, $\overline{\cal C}_{e_LA}$ is identical to Eq.~(\ref{first}). 
On the other hand, by calling $e_R$ the $E_R$ qubit with which $A$ collides, we have that $\overline{\cal C}_{e_RA}<\overline{\cal C}_{e_LA}$, regardless of $\dim(E_R)$, a behavior that can be rigorously explained as follows. For the sake of simplicity, we consider the simplest case where both the environments consist of a single qubit. By means of the decomposition in Eq.~(\ref{euler}), we prepare a general (separable) initial state of $e_R$ and $e_L$ by specifying two sets of parameters $\{\phi_j,\psi_j,\chi_j,\alpha_j\}~~(j=L,R)$. The ancilla is then let interact with $e_R$ and $e_L$, in this order. The degrees of freedom of $e_L$ are thus traced out so as to ascertain the concurrence between $A$ and $e_R$. This turns out to be
\begin{equation}
\label{third}
{\cal C}_{e_RA}=|\sin^2(\phi_R)\cos(2Jt)\sin(4Jt)|.
\end{equation}
There is no dependence on the state of $e_{L}$ in this expression, although the second collision has occurred and no average has yet been taken. However, this is quite clear in virtue of the fact that we are considering only a single interaction with each environment. After the {\it collision} with $e_L$, there is no way for $A$ to convey to $e_R$ information on the state of the left qubit. In fact, if we instead trace out $e_{R}$, we find precisely Eq.~(\ref{singleconc}). Looking for typical values, {\it i.e.} calculating averages over the tensor product of any possible unitary matrix by using the appropriate Haar measure introduced before, we get $\overline{\cal C}_{e_RA}=|\cos(2Jt)\sin(4Jt)|/2<\overline{\cal C}_{e_LA}~\forall~{Jt}$. Quantitatively, for $Jt=1$ the typical $e_R-A$ concurrence is $\simeq0.157$, which perfectly agrees with the independent numerical estimate provided in Fig.~\ref{newplot} {\bf (a)} (dashed line). The ordering relation involving $\overline{\cal C}_{e_RA}$ and $\overline{\cal C}_{e_LA}$ is preserved as the number of qubits in $E_R$ grows, as shown in Fig.~\ref{newplot} {\bf (a)}, where the difference $\overline{\cal C}_{e_LA}-\overline{\cal C}_{e_RA}$ appear to increase with $\dim(E_R)$. 
This is a clear effect of the profound differences, stemming from the possibility to achieve multipartite entanglement, between the present situation and the single-environment case addressed in Sec.~\ref{onereg}. For $\dim(E_R)>1$,
one cannot exclude that the random unitaries taken in order to prepare $E_R$ set some genuine multipartite entanglement among its elements. Consequently, the entanglement set by the dynamics studied here and shared  by $e_R$ and $A$ would be bound to follow many-body entanglement monogamy relations~\cite{monogamy}, which lower its amount as $E_R$ grows in dimension. This is also the reason behind the behavior shown in Fig.~\ref{newplot} {\bf (b)}, where by inverting the order of the ancilla-environment interactions we lose the invariance of the $e_L-A$ concurrence, an effect occurring for precisely the same reasons explained above. On the other hand, as before, the second interaction always sets the lowest degree of entanglement. It is worth remarking that Fig.~\ref{newplot} is the result of an independent numerical calculation that, although follows the formal recipe highlighted here, does not rely on the analytic result in Eq.~(\ref{third}).

\subsection{Entanglement-Structure Transitions}

The explicit introduction of multipartite entanglement in the process described until now directs us towards the main point of this study. Through the proper tuning of the dynamical and kinematic properties of the system, we shall demonstrate the ability to manipulate the sharing structure of the typical multipartite entanglement set among $A$, $e_L$ and $e_R$. For the remaining part of our study, we find it computationally more convenient to abandon concurrence for negativity~\cite{NEG,NEGmeas}. This is an entanglement measure devised from the Peres-Horodecki criterion on positivity of partial transposition (PPT). It is a necessary and sufficient condition for $2 \times 2$, $2 \times 3$ and $\infty\times\infty$ systems, and is a sufficient condition for an arbitrary system. For a bipartite case we define the negativity as~\cite{NEGmeas}
\begin{equation}
{\cal N}_{bi}=\max [0, -2 \lambda_{neg}]
\end{equation}
where $\lambda_{neg}$ is the single negative eigenvalue arising after the partial transposition of the density matrix. The convenience in using negativity stems for its straightforward generalization to the multipartite scenario: we simply have to consider all the possible bipartite splits in a given system and the geometric average of the negativity associated with any of them. This construction ensures that our measure remains an entanglement monotone~\cite{trineg}. As we are mostly interested in at most three qubits, we concentrate on {\it tripartite negativity}, which in our case reads
\begin{equation}
\label{trineg}
{\cal N}_{tri}= [{\cal N}_{A|e_Le_R}{\cal N}_{e_L|Ae_R3}{\cal N}_{e_R|Ae_L}]^{\frac{1}{3}}
\end{equation}
where 
$A|e_Le_R$, for instance, indicates the bipartition of qubit $A$ against the group of qubits $e_L$ and $e_R$.

In light of the analysis performed in Sec.~\ref{quantan} one can clearly conclude that, if tripartite entanglement is set within the $e_L-A-e_R$ system, this has to be W-class~\cite{threequbits}. In fact, we find non-zero entanglement in any bipartition obtained by tracing out one of the components of the overall system. In Fig.~\ref{newplot}, we have shown the entanglement between the ancilla and each environmental qubit involved in the collisions. Moreover, we have also checked that the $e_L-e_R$ bipartition turns out to be inseparable in each of the situations addressed there. This is a property not shared by the GHZ-class~\cite{ghz}, where by tracing one qubit out of a tripartite register, one gets separable reduced states. In order to check that W-class entanglement is indeed in order here, we have utilized the {witness}~\cite{acin,witness}
\begin{equation}
\label{witness}
\hat{\mathcal{W}}=\frac{3}{4}\openone - \ket{\text{GHZ}}\bra{\text{GHZ}}
\end{equation}
with $\ket{\text{GHZ}}=(\ket{000}+\ket{111})/\sqrt{2}$ a three-qubit GHZ state and $3/4$ the squared maximum overlap between a GHZ state and a W-class state. Eq.~(\ref{witness}) is a valid GHZ witness: a negative expectation value of $\hat{\cal W}$ successfully reveals GHZ-class entanglement. On the other hand, this construction guarantees that a positive expectation value is achieved for W-class states~\cite{acin}. By generating a sample of $1000$ random environmental states and taking $\dim(E_R)$ up to 4 (so as to match the situation depicted in Fig.~\ref{newplot}), we have numerically verified that, as far as an isotropic Heisenberg model is used to model the collisions, $\langle\hat{\cal W}\rangle>0$ for every single preparation of the environments and regardless of $\dim(E_R)$. The calculation has been performed by maximizing the overlap between the sample states and $\ket{\text{GHZ}}$ over the tensor product of three local rotations, each acting on $A$, $e_L$ and $e_R$ respectively. Moreover, the tripartite negativity defined in Eq.~(\ref{trineg}) turns out to be non-zero for any initial preparation of our statistical sample, and so is the typical value obtained by averaging over it, which guarantees genuine tripartite entangled nature of the $e_L-A-e_R$ states with strong evidence of their W-class nature. 

The task of this Section is to show that the typical entanglement-sharing structure of such tripartite state can be abruptly modified by biasing the coupling Hamiltonian towards a specific spin non-preserving model. We thus recast Eq.~(\ref{modelH}) into $(j=L,R)$
\begin{equation}
\label{biased}
\hat{H}_{b,j}=J[\sigma_{1,A} \otimes \sigma_{1,e_j} + \lambda\sum^3_{i=2}\sigma_{i,A} \otimes \sigma_{i,e_j}],
\end{equation}
where we have clearly taken $J_{2,3}=\lambda{J}~(\lambda\in\mathbb{R})$ and $J_1=J$. The reasons behind this choice are best explained by means of the following analysis. For the sake of argument we focus on single-qubit environments and take qubits $A,e_L$ and $e_R$ as prepared in their respective ground state. By considering an $e_L-A$ collision followed by an $e_R-A$ one, both ruled by Eq.~(\ref{biased}), we get
\begin{equation}
\label{plesch}
\begin{aligned}
&\ket{\psi}_{e_LAe_R}=e^{-2 i \lambda Jt } \cos (Jt-\lambda Jt
)\ket{0}_{e_L}\![\cos (Jt-\lambda Jt )\ket{00}\\
&-i\sin (Jt-\lambda Jt)\ket{11}]_{Ae_R}\!-\!\sin
(Jt-\lambda Jt)\ket{1}_{e_L}\otimes\\
&[\sin (\lambda Jt +Jt)\ket{01}
-i \cos (\lambda Jt +Jt)\ket{10}]_{Ae_R}.
\end{aligned}
\end{equation}

\begin{figure}[t]
\centerline{\psfig{figure=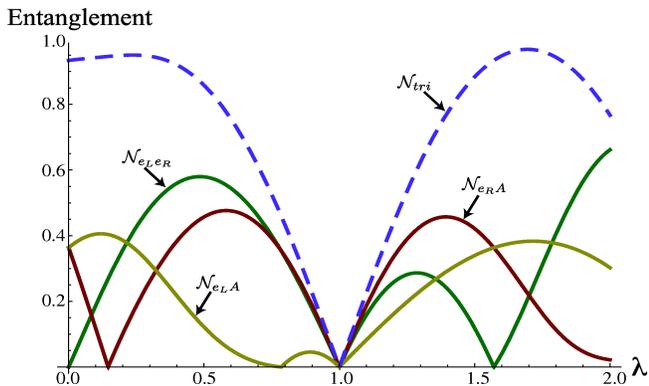,width=8.5cm,height=5cm}}
\caption{(Color online). Bipartite and tripartite negativity resulting from a bilocal interaction model ruled by Eq.~(\ref{biased}), for single-qubit environments and $\ket{000}_{e_LAe_R}$ initial state. The full disappearance of entanglement at $\lambda=1$ occurs in virtue of this specific choice of initial state. At $\lambda=0$ we have a point of entanglement-sharing change. Other three such points are visibile within the shown range of $\lambda$'s.}
\label{teoria}
\end{figure}

\begin{figure*}[ht]
{\bf (a)}\hskip7cm{\bf (b)}\hskip7cm{\bf (c)} 
\center{\psfig{figure=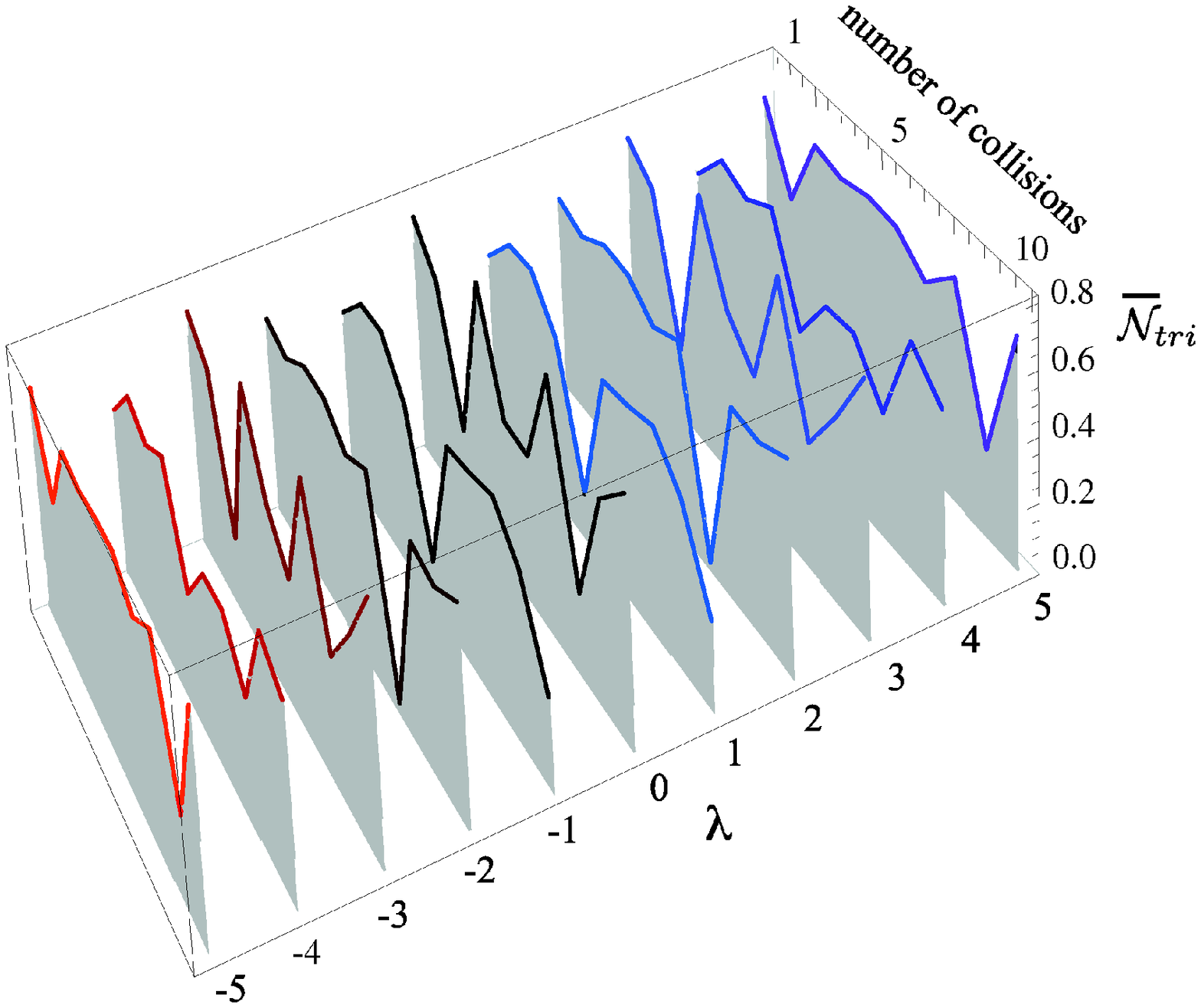,width=5.5cm,height=5.8cm}~~\psfig{figure=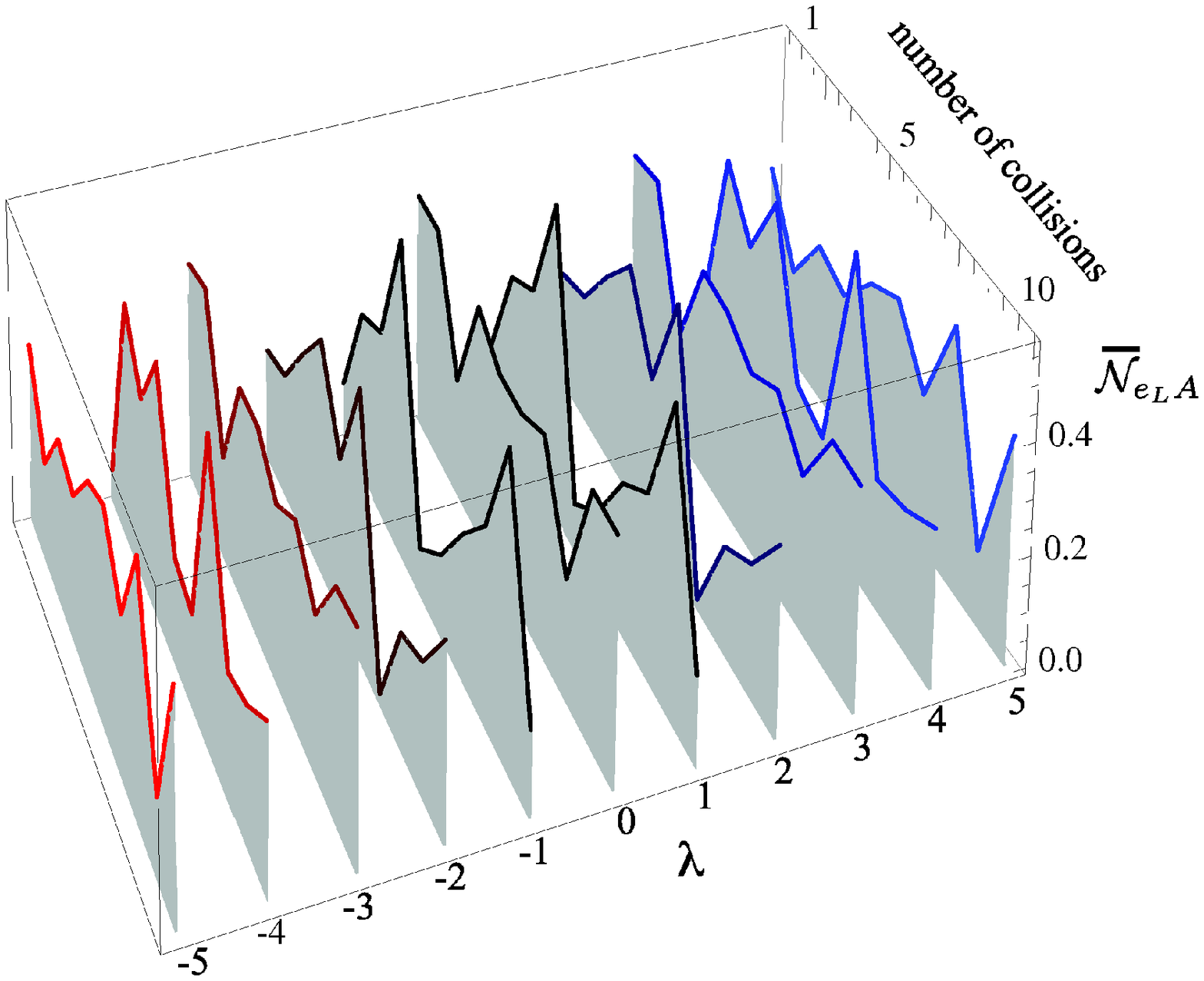,width=6.0cm,height=5.3cm}~~\psfig{figure=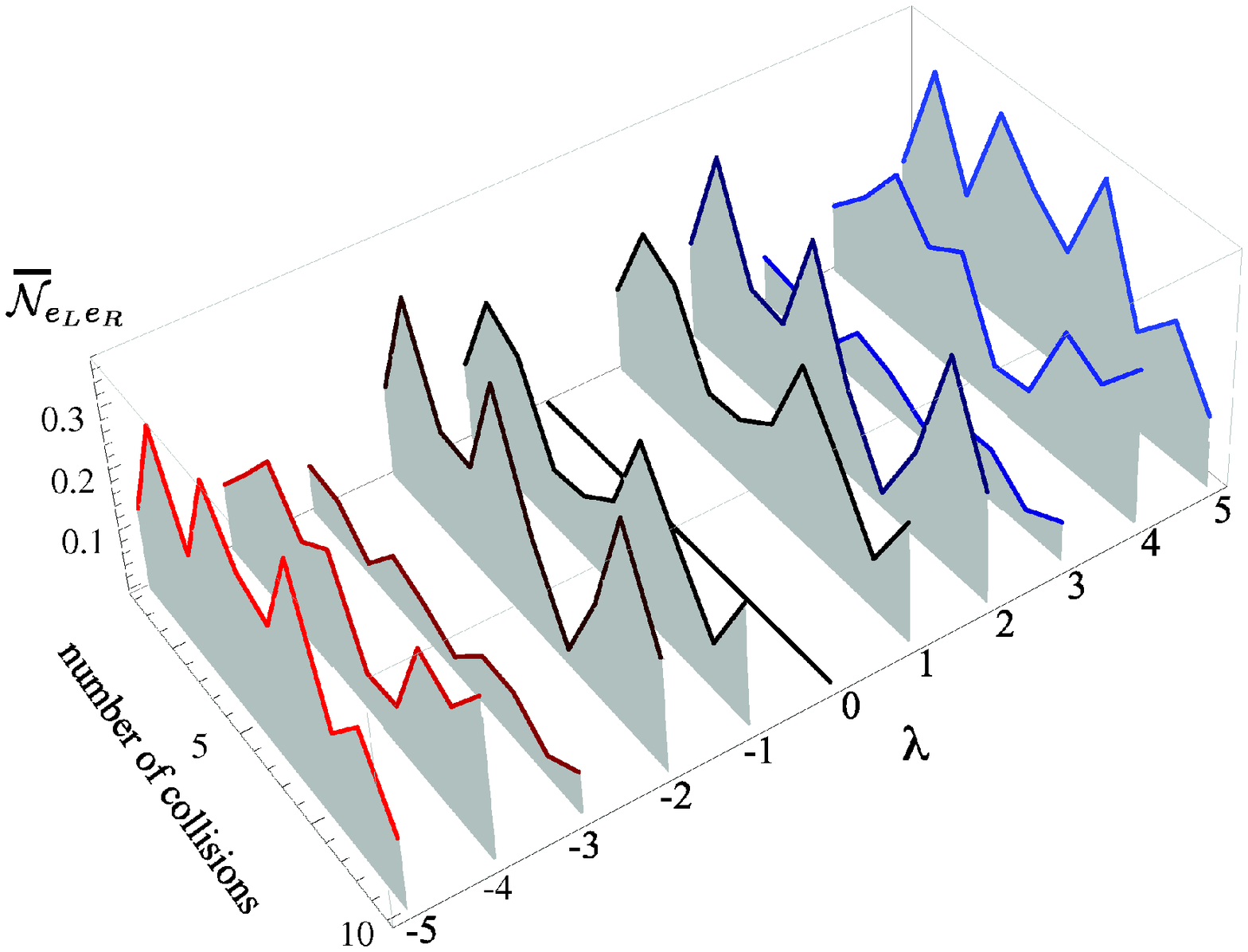,width=5.5cm,height=5.3cm}}
\caption{(Color online). {\bf (a)} Tripartite negativity of the $E_L-A-E_R$ system against $\lambda\in[-5,5]$ and the number of collisions, for a dimensionless interaction strength $Jt= 1$ in an anisotropic Heisenberg coupling involving single-qubit environments. {\bf (b)} Bipartite negativity for the $A-E_j$ qubit-pair, regardless of $j=L,R$. {\bf (c)} Bipartite negativity for the $E_L-E_R$ system. For $\lambda=0$ there is no entanglement in this bipartition, while the ancilla-environment ones are inseparable.}
\label{singleregisters}
\end{figure*}

\begin{figure}[b]
\centerline{\psfig{figure=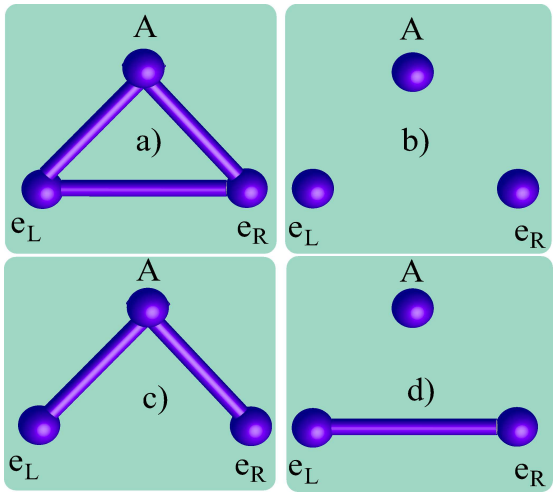,width=7.3cm,height=6.5cm}}
\caption{(Color online). Entanglement graphs for the tripartite setting. Each sphere is a qubit, each bond represents a non-separable state. {\bf (a)}  W-class states having no separable bipartition. {\bf (b)} Fully separable and GHZ-class states, with no residual bipartite entanglement. {\bf (c)} Entangled-class with bipartition-dependent residual entanglement. If the state contains genuine tripartite entanglement, it can be enumerated in the GHZ class. Systems $e_L$ and $e_R$ share classical correlations. {\bf (d)} Biseparable states encompassing a Bell pair~\cite{NC}.}
\label{entangledgraphs}
\end{figure}

Despite its innocence, the entanglement-sharing properties of Eq.~(\ref{plesch}) are quite interesting. In Fig.~\ref{teoria} we show the bipartite and tripartite negativity against the anisotropy parameter $\lambda$, for $Jt=1$ (arbitrary choice). The behavior of the bipartite entanglement is strongly dependent on the choice of $\lambda$. In particular, at $\lambda=0$, which leaves us with $\hat{H}_{b,j}=J\sigma_{1,A} \otimes \sigma_{1,e_j}$, only two out of three possible bipartitions are inseparable: the $e_L-e_R$ subsystem remains separable regardless of the choice for $Jt$. This is easily checked by studying the partial transposed of the state resulting from $\ket{\psi}_{e_LAe_R}$ upon trace over $A$. And yet, the tripartite negativity ${\cal N}_{tri}$ is rather large at $\lambda=0$ (cfr. Fig.~\ref{teoria}, dashed line), demonstrating that we have a tripartite entangled state that, in evident contrast with the results of the previous Section, is out of the W-class.~\cite{commento1}. This qualitative feature is typical as it survives to a statistical average over random initial preparations of the environments, their dimensions and the number of $A-e_j$ interactions ($j=L,R$). This latter feature is shown in Figs.~\ref{singleregisters}, where we plot the average bipartite and tripartite negativity against $\lambda\in[-5,5]$ and the number of collisions. Fig.~\ref{singleregisters} {\bf (c)} shows that at $\lambda=0$ there is no quantum correlations shared by the two environments, even in this statistically typical scenario. 

\begin{figure*}[t]
{\bf (a)}\hskip5cm{\bf (b)}\hskip5cm{\bf (c)}
\center{\psfig{figure=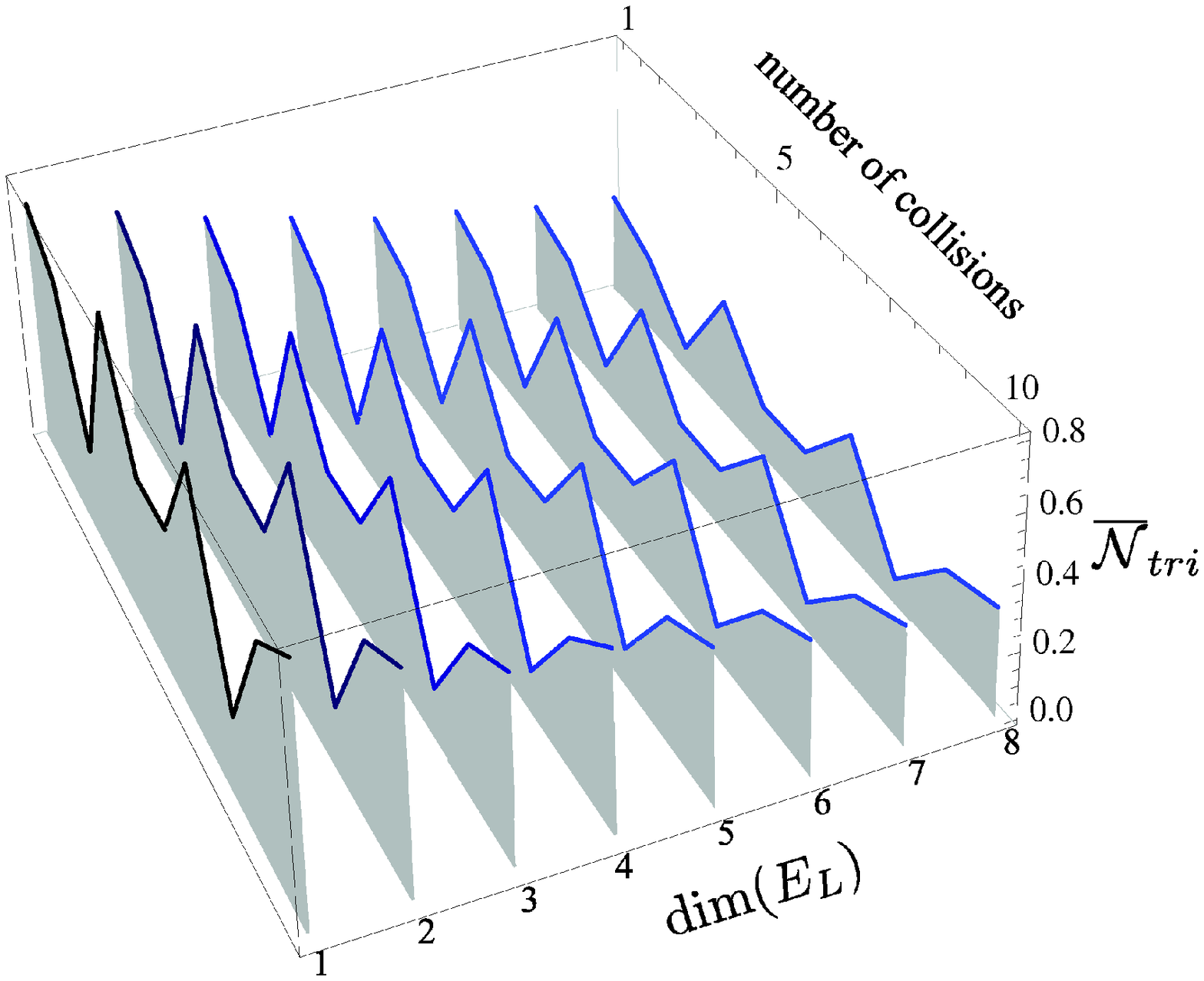,width=5.5cm,height=5cm}\,\psfig{figure=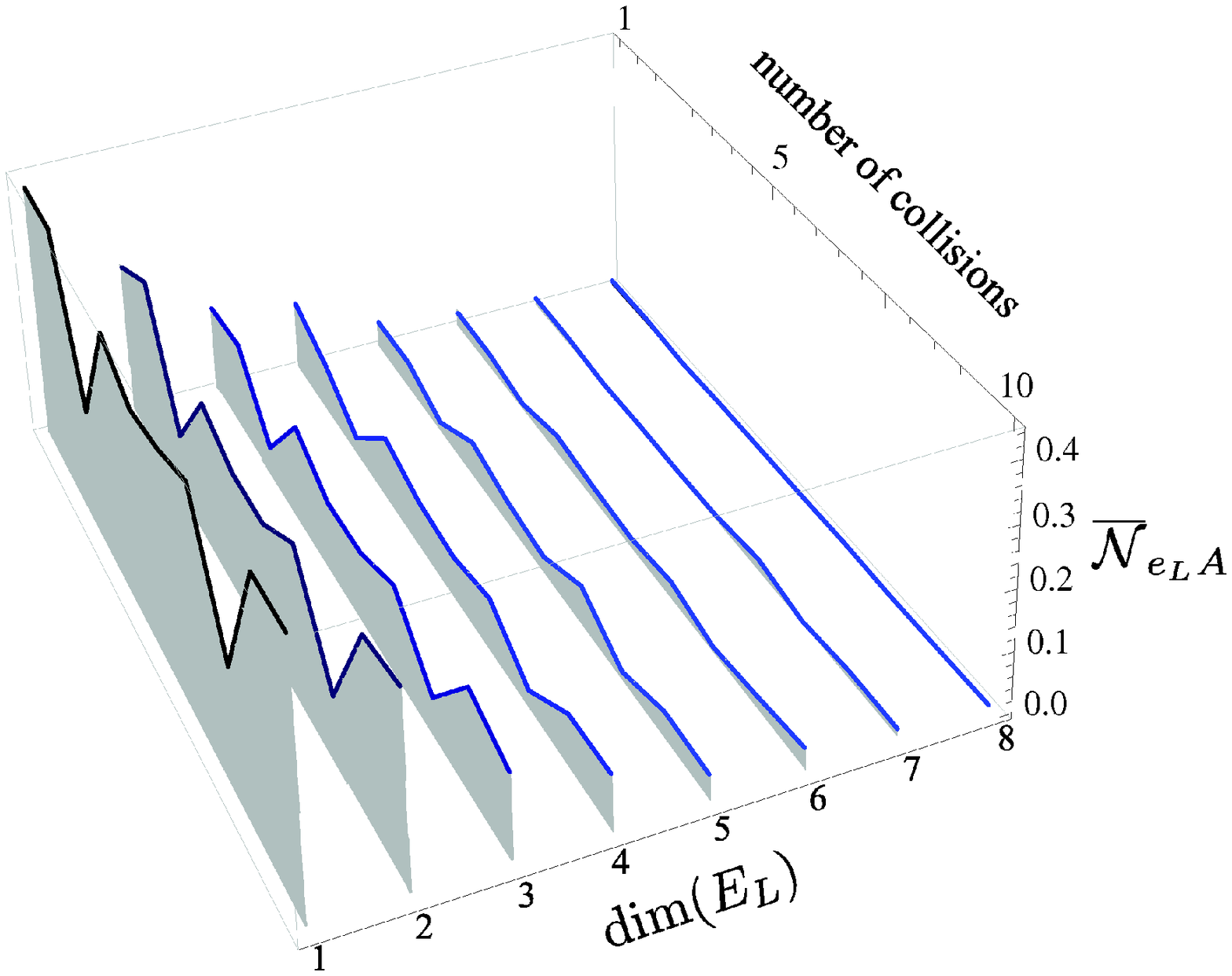,width=5.5cm,height=5cm}\,
\psfig{figure=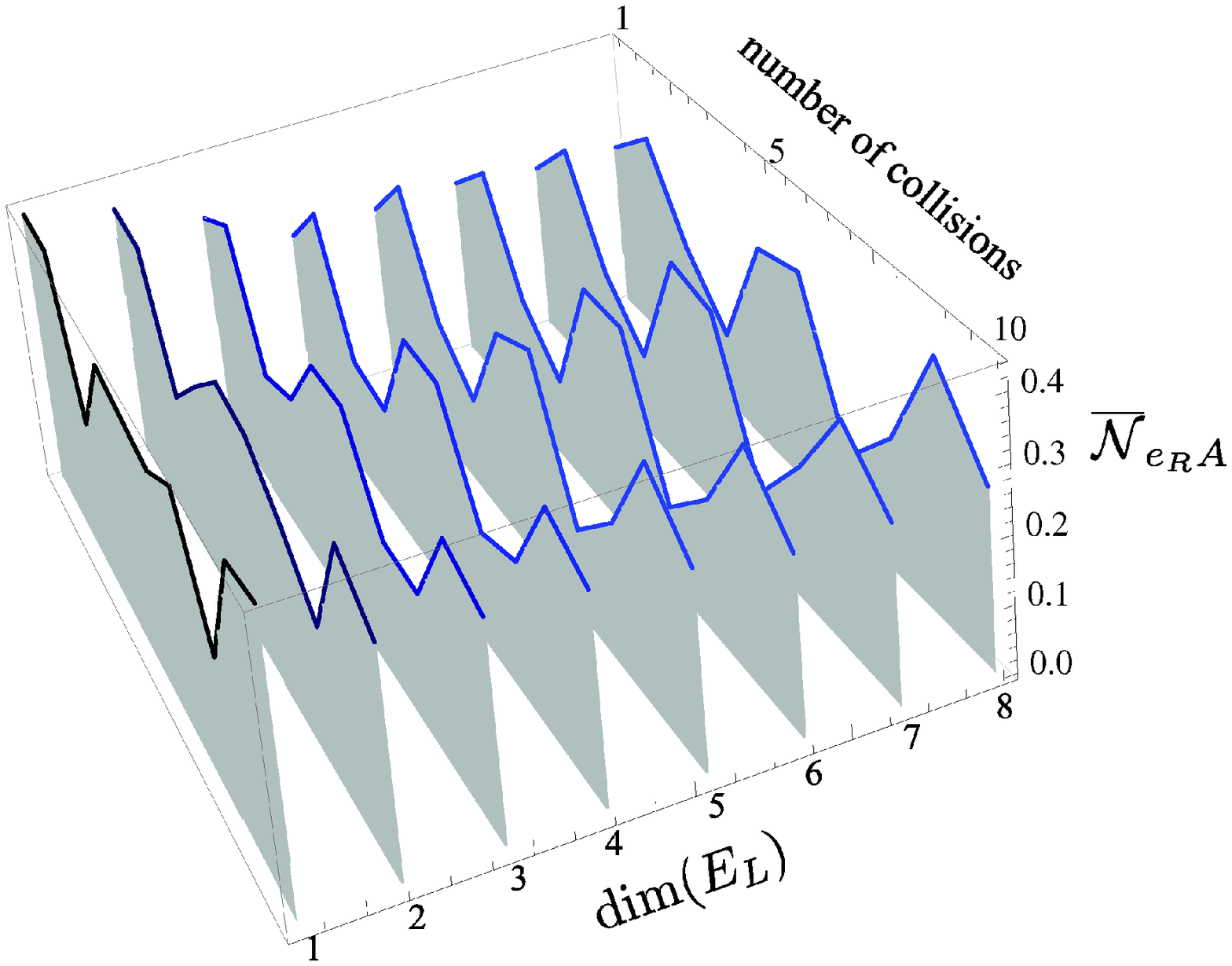,width=5.5cm,height=5cm}}
\caption{(Color online). Entanglement obtained for an increasingly-sized $E_L$ environment (which comprises of up to eight qubits) and a single-qubit $E_R$. We set $\lambda=0$ in Eq.~(\ref{plesch}) and the rescaled interaction time $Jt= 1$. {\bf (a)} Tripartite negativity. {\bf (b)} Bipartite $e_L-A$ entanglement against the dimension of $E_L$ and the number of collisions. {\bf (c)} Same as panel {\bf (b)} but for the single-qubit environment $E_R$.}
\label{biasedregisters}
\end{figure*} 

How can we understand this result and relate it to the claimed change in multipartite entanglement structure? We refer to the studies conducted by Plesch and Bu\v{z}ek~\cite{buzek} on the classification of multiparticle quantum correlations  via entangled graphs. Following their lines, we represent a multi-qubit system with an inherent structure of shared entanglement as a connected graph. Each vertex embodies a qubit while a bond connecting two vertices represents bipartite entanglement shared by the components of the corresponding reduced state. For three qubits, four possible classes are identified, as shown in Fig.~\ref{entangledgraphs}. In particular, Eq.~(\ref{plesch}), as well as any of the states resulting from the application of random unitaries over the chosen reference state $\ket{000}_{e_LAe_R}$, have a sharing structure corresponding to the graph in Fig.~\ref{entangledgraphs} {\bf (c)}, panel {\bf (a)} being representative of W-class. As it is argued in Ref.~\cite{buzek}, by following this graph-based classification of multipartite entanglement, a state showing tripartite correlations with the {\it two-way} residual entanglement structure of panel {\bf (c)} is GHZ-class. Moreover, a remark is due: such a two-way entanglement-sharing structure is possible, for pure states, only if there exist classical correlation between the separable qubits ($e_L$ and $e_R$ in our case). This property is readily confirmed in Eq.~(\ref{plesch}) and any random element of our statistical sample. In fact, we find that 
\begin{equation}
\varrho_{e_Re_L} \neq \varrho_{e_L} \otimes \varrho_{e_R}
\end{equation}
where $\varrho_{e_{L,R}}$ denotes the reduced density matrix of qubit $e_{L,R}$ after (random) preparation, unitary evolution and appropriate partial traces~\cite{commento2}. We can thus rightfully claim for the anticipated {entanglement transition} induced by the tuning of the effective {\it order parameter} embodied by $\lambda$. The naturally generated W-class state arising from the use of an unbiased Heisenberg coupling is non-trivially changed into a GHZ-type entanglement characterized by a two-way sharing structure. The transition is typical in the sense explained so far: given {\it any} pure initial environmental state, we are able to create either type of tripartite entanglement structures simply by tuning the interaction properties.

\begin{figure}[b]
{\bf (a)}\hskip3cm{\bf (b)}
\centerline{\psfig{figure=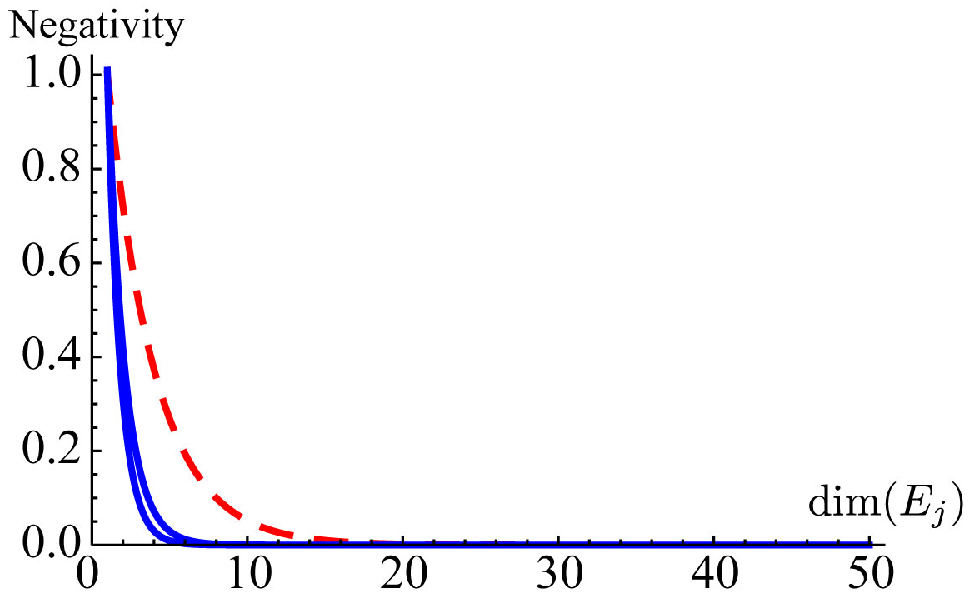,width=4.8cm,height=3.2cm}\psfig{figure=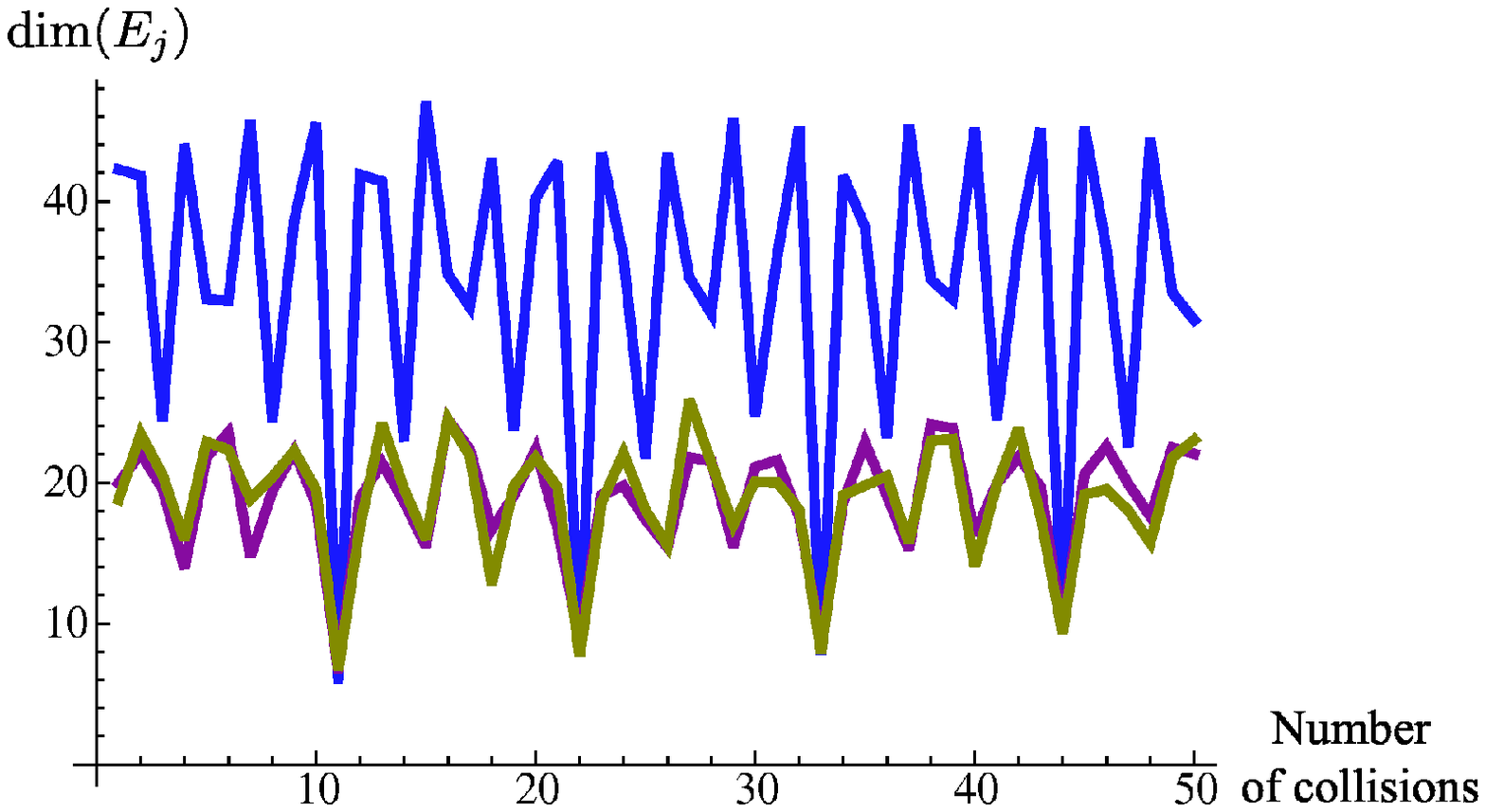,width=4.8cm,height=3.2cm}}
\caption{(Color online).  {\bf (a)} Negativity against the environments' dimension $\dim(E_j)~(j=L,R)$ for $\lambda=0$, 18 pairs of collisions and $Jt=1$. The dashed line shows the tripartite negativity, while the solid lines are for the entanglement within each bipartite subsystem involving only one environment. The $e_R-e_L$ entanglement is obviously null. Each curve is rescaled with respect to its maximum value. {\bf (b)}: We plot the number of  qubits required in the remote environments in order to make the tripartite and bipartite negativity smaller than $10^{-6}$ when a fixed number of collisions is taken. The topmost curve is for ${\cal N}_{tri}$ while the other two curves, which are almost superimposed, show the cases associated with ${\cal N}_{e_jA}$.}
\label{fittato}
\end{figure}

As anticipated, we also allow the ancilla to interact multiple times, in sequence, with each environment (each time with the very same $e_{L,R}$ qubit). Multiple interactions, however, do not appear to have any qualitative relevance on the generated entanglement structure, although there is an evident quantitative effect on the amount of entanglement shared both in bipartite and tripartite sense. In Fig.~\ref{singleregisters} {\bf (a)} and {\bf (b)} we plot the typical tripartite and bipartite entanglement of the $A-e_L$ system and it is easily verified that the same behavior is found for the $A-e_R$ system. For the whole considered range of values for $\lambda$ the bipartite negativity is always non zero, signaling that this interaction model guarantees the generation of entanglement between the ancilla and an environmental qubit even under the ``minimum-control" conditions where one is unable to stop the ancilla-environment interactions after a desired number of collisions. Moreover, as claimed above, at $\lambda=0$ the $e_L-e_R$ system is indeed separable, regardless of the number of considered collisions (see the $\overline{\cal N}_{e_Le_R}=0$ line in Fig.~\ref{singleregisters} {\bf (c)}).

It is very informative to study the effects that increasing the size of the environments has on the degree and type of shared entanglement for $\lambda=0$. We first take $\dim(E_L)\ge1$ and leave $E_R$ with a single qubit. The scheme of interaction in the usual one, with $e_L$ being struck first. The features associated with the resulting entanglement sharing are shown in Figs.~\ref{biasedregisters}. In panel {\bf (a)} we plot the tripartite negativity set within the state of the interacting qubits after tracing out the non-interacting elements from the $E_L$ environment. We find that $\overline{\cal N}_{tri}$ decreases with the dimension of $E_L$, which is a physically meaningful result in light of monogamy relations that the system should adhere to~\cite{monogamy}. A similar behavior is found for the bipartite $e_L-A$ entanglement, as shown in Fig~\ref{biasedregisters} {\bf (b)}, which can be easily interpreted in light of the discussion in Sec.~\ref{quantan}. Finally, panel {\bf (c)} shows the $e_R-A$ entanglement, which exhibits a constant trend that is consistent with our previous results on the isotropic Heisenberg Hamiltonian. We omit showing the $e_R-e_L$ entanglement as this is exactly null, in these conditions. The entanglement structure remains of the two-way GHZ class~\cite{buzek} although for $\dim(E_L)\gtrsim{8}$, ${\cal N}_{e_LA}\simeq{0}$. This implies the disappearance of the leftmost bond in Fig.~\ref{entangledgraphs} {\bf (c)} but no further change, as genuine tripartite entanglement is still present in the $e_L-A-e_R$ system (see Fig.~\ref{biasedregisters} {\bf (a)}). 

The above analysis covers only partially the issue of increasing-sized environments, as it addresses an asymmetric situation. Another interesting effect arises if $E_R$ is let grow in size. In fact, the rate at which bipartite quantum correlations decrease in the case considered above suggests that, by increasing $\dim(E_R)$, all residual entanglement would die off, eventually, leaving us with a sharing structure that reminds more of a typical GHZ state. In other words, we are looking for a configuration allowing an additional entanglement transition from a graph of class {\bf (c)} to one of class {\bf (b)} in Fig.~\ref{entangledgraphs}.  This is indeed possible. We consider $\dim(E_L)=\dim(E_R)$ and investigate the behavior of both tripartite and bipartite negativity against the number of environmental qubits and the number of interactions with the ancilla. A typical result, achieved by considering $18$ pairs of collisions, is given in Fig.~\ref{fittato} {\bf (a)} where it is evident that, by considering larger environments, we are able to keep the tripartite entanglement non-zero using repeated collisions between $A$ and the environments. Any two-qubit quantum correlation, on the other hand, disappears. The curves are the best-fits to the discrete numerical values obtained via our simulations. We have found that the functional form ${\cal A}{e}^{-{\cal B}(\dim(E_j)-1)}$ (with ${\cal A},{\cal B}\in\mathbb{R}$) gives an excellent agreement with the numerical results.  Fig.~(\ref{fittato}) {\bf (b)} shows the number of qubits necessary in $E_j~(j=R,L)$ in order to get negativities smaller than $10^{-6}$ (arbitrarily chosen) at a fixed number of collisions. It is clearly seen that the environmental dimension required to get entanglement lower than the chosen threshold is almost always much larger for the tripartite negativity than for the bipartite ones, thus strengthening our conclusions: we have been able to {dynamically} drive the system state towards {typical} GHZ-class entanglement, therefore realizing yet another sharing-structure transition. 

\section{conclusions}
\label{conclusions}
Using a random-matrix approach, we have shown that the typical entanglement properties of two remote qubit environments, indirectly communicating via the mediation operated by a shuttling ancilla, can be effectively manipulated both quantitatively and qualitatively. Bipartite as well as genuinely multipartite entanglement can be faithfully generated and shown to have intriguing statistically averaged properties of resilience against arbitrary pure preparations of the environments, their dimension and the number of interactions that each has with the shuttle. Interestingly, by means of a simple tuning of the details of the interaction, we can achieve transitions in the entanglement-sharing structure from W to two-way GHZ-class~\cite{buzek}. We have studied the effects that an increasing dimension of the environments have in such transitions. The use of random matrices has emerged, here, as a valuable tool for the agile handling of the numerical simulations required for the purposes of our study, enlarging the range of applicability of the technique to the context of multipartite entanglement generation under unfavorable conditions. This is a central problem in current quantum information and computation problems, where we hope that our results will trigger further interest. We are currently investigating strategies, developed well along the lines of this paper, to study the interplay between quantum correlations and  thermodynamical properties of a multipartite system.  

\acknowledgments 

MP thanks Dr. M. Guta for interesting discussions on the topic of entanglement in random-matrix settings. This work has been supported by PRIN 2006 ``Quantum noise in mesoscopic systems", EUROTECH, DEL, the British Council/MIUR British-Italian Partnership Programme 2007-2008 and the UK EPSRC (EP/G004579/1). SC gratefully acknowledges the hospitality of the DSFSA, Universita' degli Studi di Palermo, where part of this work has been performed.

\renewcommand{\theequation}{A-\arabic{equation}}
\setcounter{equation}{0}
\section*{APPENDIX}
\subsection*{Construction of Random Unitary Matrices}
\label{Hurwitz}
Here we briefly outline the recipe to generate a random unitary matrix. The algorithm has been extensively used in recent works~\cite{kus1991,kus1994, kus1996,kus1998,Weinstein2005}. The parameterisation is based on the original work presented by Hurwitz in 1897~\cite{hurwitz}. Any unitary matrix, $U_r$, of dimension $n$, can be decomposed as
\begin{equation}
U_r=e^{i \alpha} E_1 E_2 \dots E_{n-1}
\end{equation}
where $E_{i}$ is an $n\times{n}$ matrix. Matrices $E_{i}$'s are readily constructed using products of proper rotation matrices $R^{(i,j)}(\phi_{i\,j},\psi_{i\,j},\chi_{i\,j})$, each depending on the respective set of Euler's angles $\{\phi_{i\,j},\psi_{i\,j},\chi_{i\,j}\}$ as follows
\begin{equation}
\begin{aligned}
E_1 &= R^{(1,2)}(\phi_{1\,2},\psi_{1\,2},\chi_{1\,2}),\\
E_2 &= R^{(2,3)}(\phi_{2\,3},\psi_{2\,3}, 0) R^{(1,3)}(\phi_{1\,3},\psi_{1\,3},\chi_{1\,3}),\\
E_3 &= R^{(3,4)}(\phi_{3\,4},\psi_{3\,4},0) R^{(2,4)}(\phi_{2\,4},\psi_{2\,4},0),\\
          &\times{R}^{(1,4)}(\phi_{1\,4},\psi_{1\,4},\chi_{1\,4}),\\
&\vdots \\
E_{n-1} &= R^{(n-1,n)}(\phi_{n-1\,n},\psi_{n-1\,n},0)\times\dots \\
               &\times{R}^{(1,n)}(\phi_{1\,n},\psi_{1\,n},\chi_{1\,n}).
\end{aligned}
\end{equation}
The matrix elements are taken as
\begin{equation}
\begin{aligned}
R^{(i,j)}_{k,k} &= 1~~(\text{for}~k \neq i,j),\\
R^{(i,j)}_{i,i} &= e^{i\psi}\cos\phi,~~R^{(i,j)}_{i,j} = e^{i\chi}\sin\phi,\\
R^{(i,j)}_{j,i} &= -e^{-i\chi}\sin\phi~~R^{(i,j)}_{i,j} = e^{-i\psi}\cos\phi,\\
\end{aligned}
\end{equation}
and zero otherwise. The angles are drawn from the ranges
$0 \leq \phi_{i\,j} \leq {\pi}/{2},~0 \leq \psi_{i\,j} \leq 2\pi,~0 \leq \chi_{i\,j} \leq 2\pi$ and $0\le\alpha\le{2\pi},$ uniformly with respect to the corresponding (and properly normalized) Haar measure~\cite{kus1994}.


\begin{thebibliography}{99}
\bibitem{referee2} It is sufficient to consider the following simple example to see that the dimension of a multipartite environment prepared in an entangled state can affect the bipartite quantum correlations between an ancilla $A$ and the qubit $e$ of the environment it interacts with. For the sake of argument, let us assume the $e-A$ interaction is embodied, in this example, by a controlled-NOT gate~\cite{NC}. While $A$ is prepared in a superposition state $\ket{+}_A$ such that $\hat{\sigma}_x\ket{+}_A=\ket{+}_A$, the environment is off-line prepared in an $N$-qubit W state $(1/\sqrt{N})\sum^N_{j=1}\ket{0_10_2\cdot\cdot{1}_j\cdot\cdot{0}_N}$. For definiteness, let us take $e=1$. It is straightforward to see that upon $e-A$ interaction and trace over the $N-1$ environmental qubits not involved in such a dynamics, the bipartite entanglement shared by $e$ and the ancilla does indeed depend on $N$. Quantitatively, ${\cal C}_{eA}=1/3$ for $N=3$ while it is $1/2$ for $N=4$. Clearly, this is simply an example, which is nevertheless significant enough to illustrate the potential dependence of ${\cal C}_{eA}$ on ${\text{dim}}(E)$ under the assumptions of our work.
\bibitem{commento1} The fact that, at $\lambda=1$, the state is separable both in the bipartite and tripartite sense should not surprise in light of the choice of initial state $\ket{000}_{e_LAe_R}$ and the spin-preserving nature of the isotropic Heisenberg Hamiltonian.
\bibitem{commento2} Further confirmation on the change in entanglement structure is seen though the use of the entanglement witness in Eq.~(\ref{witness}), whose expectation value becomes negative.
\end{thebibliography}
\end{document}